\documentclass[12 pt,twoside,aps,prd,amsmath,amssymb,
tightenlines,showpacs,showkeys,eqsecnum]{revtex4}

\usepackage{epsfig}
\usepackage{graphicx}
\usepackage{mathptmx}
\usepackage{bm}

\newcommand {\n}{¹}

\newcommand {\ve}{\varepsilon}

\newcommand {\cG}{\cal G}
\newcommand {\cD}{\cal D}
\newcommand {\bg}{\bar \gamma}
\newcommand {\G}{\Gamma}
\newcommand {\bp}{\bar \psi}

\def\myfigure #1#2#3#4
{\begin{figure}[ht]\begin{center}
\includegraphics[width=#2 \textwidth]{#1.eps}
\parbox[t]{#4\textwidth}{\caption{#3}\label{#1}}
\end{center}\end{figure}}

\def \myfigures #1#2#3#4#5#6#7#8
{\begin{figure}[ht]
    \begin{center}
        \includegraphics[width=#2 \textwidth]{#1.eps}
        \hfill
        \includegraphics[width=#6 \textwidth]{#5.eps}
        \parbox[t]{#4\textwidth}{\caption {#3}\label{#1}}
        \hfill
        \parbox[t]{#8\textwidth}{\caption {#7}\label{#5}}
    \end{center}
\end{figure} }

\begin{document}

\baselineskip -24pt
\title{Nonlinear Spinor field in
isotropic space-time and dark energy models}
\author{Bijan Saha}
\affiliation{Laboratory of Information Technologies\\
Joint Institute for Nuclear Research\\
141980 Dubna, Moscow region, Russia} \email{bijan@jinr.ru}
\homepage{http://spinor.bijansaha.ru}

\begin{abstract}

Within the scope of isotropic FRW cosmological model the role of
nonlinear spinor field in the evolution of the Universe is studied.
It is found that unlike in anisotropic cosmological models in the
present case the spinor field does not possess nontrivial
non-diagonal components of energy-momentum tensor, consequently does
not impose any additional restrictions on the components of the
spinor field or metric function. The spinor description of different
matter was given and evolution of the Universe corresponding to
these sources is illustrated. In the framework of a three fluid
system the utility of spinor description of matter is established.
\end{abstract}

\keywords{Spinor field, late time acceleration, oscillatory
solution, quintom, three fluid}

\pacs{98.80.Cq}

\maketitle

\bigskip

\section{Introduction}

Nonlinear self-couplings of the spinor fields may arise as a
consequence of the geometrical structure of the space-time and, more
precisely, because of the existence of torsion. As early as 1938,
Iwanenko \cite{ivanenko1} showed that a relativistic theory imposes
in some cases a fourth-order self-coupling. This theory was further
developed in \cite{ivanenko2,ivanenko3,rodichev}. The influence of
nonlinear (fourth-order) terms in the Lagrangian of some classical
relativistic field theories was investigated in \cite{soler1}. In
case of spinor field, stable localized configurations with a lowest
energy state are shown to exist always for positive values of the
coupling constant. As the self-action is of spin-spin type, it
allows the assignment of a dynamical role to the spin and offers a
clue about the origin of the nonlinearities. This question was
further clarified in some important papers by Utiyama, Kibble, and
Sciama ~\cite{utiyama,kibble,sciama}. Particle-like solutions of
classical spinor field equations were obtained in
\cite{Finkle1,Finkle2,hb1}. Stability of optical gap solitons, i.e.
localized solutions of spinor-like system, is analyzed in
\cite{Barashenkov}. A nonlinear spinor field, suggested by the
symmetric coupling between nucleons, muons, and leptons, was
investigated in \cite{Finkle1} in the classical approximation. A
classical spinor field defined by a variational principle on a
Lagrangian with quadratic Dirac and quartic Fermi terms was
investigated in \cite{Finkle2}. In the simplest scheme, the
self-action is of pseudovector type, but it can be shown that one
can also get a scalar coupling ~\cite{soler}. An excellent review of
the problem may be found in~\cite{hehl}. Nonlinear quantum Dirac
fields were used by Heisenberg~\cite{hb1,hb2} in his ambitious
unified theory of elementary particles. They are presently the
object of renewed interest since the widely known paper by Gross and
Neveu~\cite{gross} where the two-dimensional massless fermion field
theories with quartic interaction were studied. Nonlinear spinor
field within the scope of static plane-symmetric model of
gravitational field was studied in
\cite{Saha2002JTMCP,Saha2004CzJP,Saha2005IJTP,SahaBook}. Recently a
variational method for studying the evolution of solitary wave
solution of nonlinear Dirac equation was developed in \cite{Khare}.

But thanks to its ability to describe different stages of evolution
\cite{henneaux,ochs,Saha1997GRG,Saha1997JMP,Saha2001PRD,greene,Saha2004aPRD,
Saha2004bPRD,ribas,Saha2006ECAA,Saha2006GnC,Saha2006PRD,Saha2007RRP,souza,kremer}
as well as simulate different characteristic of matter from perfect
fluid to phantom matter nonlinear spinor field is now extensively
exploited in cosmology
\cite{Krechet,Saha2010CEJP,Saha2010RRP,Saha2011APSS,Saha2012IJTP}.

But some recent studies showed that the presence of non-trivial
non-diagonal components of the energy-momentum tensor of the spinor
field plays very crucial role in the evolution of both spinor field
and the metric functions
\cite{FabIJTP,ELKO,F1,F2,Saha2014IJTP,Saha2015APSS,Saha2015EPJP,Saha2015CJP,Saha2015CnJP,Saha2016IJTP,Saha2016EPJP}.
Unlike in anisotropic cosmological models the non-diagonal
components of the energy-momentum tensor of the spinor field in the
isotropic FRW space-time are trivial. Moreover, the FRW model gives
a surprisingly accurate picture of the present day Universe. Hence
in this paper we study the role of nonlinear spinor field in the
evolution of an isotropic and homogeneous FRW Universe. Cosmological
models with nonlinear spinor field within scope of FRW space-time
was studied in \cite{Saha2014IJTP,Saha2015APSS,Ribas2016}. The
purpose of this paper is to study the role of spinor field
nonlinearity in the evolution of the isotropic space-time. Beside
this we give spinor descriptions of fluid and dark energy and show
why this method is convenient to exploit to study the evolution of
the Universe.

\section{Basic equation}

Let us consider the case when the isotropic and homogeneous
space-time is filled with nonlinear spinor field. The corresponding
action can be given by
\begin{equation}
{\cal S}(g; \psi, \bp) = \int\, L \sqrt{-g} d\Omega \label{action}
\end{equation}
with
\begin{equation}
L= L_{\rm g} + L_{\rm sp}. \label{lag}
\end{equation}
Here $L_{\rm g}$ corresponds to the gravitational field
\begin{equation}
L_{\rm g} = \frac{R}{2\kappa}, \label{lgrav}
\end{equation}
where $R$ is the scalar curvature, $\kappa = 8 \pi G$ is the
Einstein's gravitational constant and $L_{\rm sp}$ is the spinor
field Lagrangian.

Let us consider the isotropic FRW space-time given by
\begin{equation}
ds^2 = dt^2 - a^{2} \left[dx^{2}\, + \,dy^{2} + \,dz^2\right],
\label{frw}
\end{equation}
with $a$ being the functions of time only.

The spinor field Lagrangian is given by

\begin{equation}
L_{\rm sp} = \frac{\imath}{2} \left[\bp \gamma^{\mu} \nabla_{\mu}
\psi- \nabla_{\mu} \bar \psi \gamma^{\mu} \psi \right] - m_{\rm sp}
\bp \psi - F. \label{lspin}
\end{equation}

We choose the nonlinear term $F$ to be the function of $K$ only,
i.e., $F = F(K)$, with $K$ taking one of the following expressions
$\{I,\,J,\,I+J,\,I-J\}$, where $I$ is the scalar bilinear invariant:
$I = S^2 = (\bp \psi)^2$ and $J$ is the pseudoscalar bilinear
invariant: $J = P^2 = (\imath \bp \gamma^5 \psi)^2$. In
\eqref{lspin} $\nabla_\mu$ is the covariant derivative of spinor
field:
\begin{equation}
\nabla_\mu \psi = \frac{\partial \psi}{\partial x^\mu} -\G_\mu \psi,
\quad \nabla_\mu \bp = \frac{\partial \bp}{\partial x^\mu} + \bp
\G_\mu, \label{covder}
\end{equation}
with $\G_\mu$ being the spinor affine connection.

Variation of \eqref{lspin} with respect to $\bp$ and $\psi$ yields
spinor field equations
\begin{subequations}
\label{speq}
\begin{eqnarray}
\imath\gamma^\mu \nabla_\mu \psi - m_{\rm sp} \psi - {\cD} \psi -
 \imath {\cG} \gamma^5 \psi &=&0, \label{speq1} \\
\imath \nabla_\mu \bp \gamma^\mu +  m_{\rm sp} \bp + {\cD}\bp +
\imath {\cG} \bp \gamma^5 &=& 0, \label{speq2}
\end{eqnarray}
\end{subequations}
where we denote ${\cD} = 2 S F_K K_I$ and ${\cG} = 2 P F_K K_J$,
with $F_K = dF/dK$, $K_I = dK/dI$ and $K_J = dK/dJ.$ In view of
\eqref{speq} the spinor field Lagrangian can be rewritten as
\begin{eqnarray}
L_{\rm sp} & = & \frac{\imath}{2} \left[\bp \gamma^{\mu}
\nabla_{\mu} \psi- \nabla_{\mu} \bar \psi \gamma^{\mu} \psi \right]
- m_{\rm sp} \bp \psi - F(I,\,J)
\nonumber \\
& = & \frac{\imath}{2} \bp [\gamma^{\mu} \nabla_{\mu} \psi - m_{\rm
sp} \psi] - \frac{\imath}{2}[\nabla_{\mu} \bar \psi \gamma^{\mu} +
m_{\rm sp} \bp] \psi
- F(I,\,J),\nonumber \\
& = & 2 F_K (I K_I + J K_J) - F = 2 K F_K - F(K). \label{lspin01}
\end{eqnarray}

Choosing the tetrad for the metric \eqref{frw} in the following way

\begin{equation}
e_0^{(0)} = 1, \quad e_1^{(1)} = a, \quad e_2^{(2)} = a, \quad
e_3^{(3)} = a. \label{tetradfrw}
\end{equation}

from
\begin{equation}
\Gamma_\mu = \frac{1}{4} \bg_{a} \gamma^\nu \partial_\mu e^{(a)}_\nu
- \frac{1}{4} \gamma_\rho \gamma^\nu \Gamma^{\rho}_{\mu\nu}.
\label{sfc}
\end{equation}
one finds the expressions for spinor affine connections as:
\begin{equation}
\G_0 = 0, \quad \G_1 = \frac{\dot a}{2} \bg^1 \bg^0, \quad \G_2 =
\frac{\dot a}{2} \bg^2 \bg^0, \quad \G_3 = \frac{\dot a}{2} \bg^3
\bg^0. \label{sac123frw}
\end{equation}

From the definition \cite{Saha2001PRD}
\begin{equation}
T_{\mu}^{\,\,\,\rho}=\frac{\imath}{4} g^{\rho\nu} \biggl(\bp
\gamma_\mu \nabla_\nu \psi + \bp \gamma_\nu \nabla_\mu \psi -
\nabla_\mu \bar \psi \gamma_\nu \psi - \nabla_\nu \bp \gamma_\mu
\psi \biggr) \,- \delta_{\mu}^{\rho} L_{\rm sp}. \label{temsp}
\end{equation}
in this case one finds the the following nontrivial components of
energy-momentum tensor of the spinor field
\begin{subequations}
\label{emtfrw}
\begin{eqnarray}
T_0^0 & = & m_{\rm sp} S + F(K), \label{emt00frw}\\
T_1^1 &=& T_2^2 = T_3^3 = F(K) - 2 K F_K, \label{emtiifrw}
\end{eqnarray}
\end{subequations}

Further, taking into account that the Einstein tensor corresponding
to the metric \eqref{frw} has only nontrivial diagonal components:

\begin{subequations}
\label{LRSBIET}
\begin{eqnarray}
G_1^1 &=& G_2^2 = G_3^3 = - \left(2\frac{\ddot a}{a} + \frac{\dot
a^2}{a^2}\right),\label{frwETii}\\
G_0^0 &=& - 3\frac{\dot a^2}{a^2}. \label{frwET00}
\end{eqnarray}
\end{subequations}

we write the complete set of Einstein equation for FRW metric
\begin{subequations}
\label{frwEn}
\begin{eqnarray}
2\frac{\ddot a}{a}  + \frac{\dot
a^2}{a^2} &=& \kappa (F(K) - 2 K F_K),\label{11frw}\\
3\frac{\dot a^2}{a^2}&=& \kappa (m_{\rm sp} S + F(K)). \label{00frw}
\end{eqnarray}
\end{subequations}
Note that one can solve \eqref{00frw} to find $a$, but to obtain the
solution that satisfies both \eqref{11frw} and \eqref{00frw} it is
worthy to combine them as follows:
\begin{equation}
\ddot a = \frac{\kappa}{6}\left(3 T_1^1 - T_0^0\right)\,a =
\frac{\kappa}{6}\left(2 F(K) - 6 K F_K -  m_{\rm sp} S \right)\, a.
\label{frw0}
\end{equation}

To solve the equation \eqref{frw0} one should know the explicit form
of spinor field nonlinearity and also the relation between the
invariant $K$ and metric function $a$. To find this relation, let us
go back to spinor field equations. In this case in view of
\eqref{covder} and \eqref{sac123frw} the spinor field equations
\eqref{speq} take the form

\begin{subequations}
\label{SF1frw}
\begin{eqnarray}
\imath \bg^0 \left(\dot \psi + \frac{3}{2}\frac{\dot a}{a}
\psi\right) - m_{\rm sp} \psi - {\cD}
\psi -  \imath {\cG} \bg^5 \psi &=&0, \label{speq1pfrw}\\
\imath \left(\dot \bp + \frac{3}{2}\frac{\dot a}{a} \bp\right)\bg^0
+ m_{\rm sp} \bp  + {\cD}  \bp + \imath {\cG}\bp \bg^5 &=& 0.
\label{speq2pfrw}
\end{eqnarray}
\end{subequations}
From \eqref{SF1frw} it can be shown that \cite{Saha2015EPJP}
\begin{equation}
K = \frac{V_0^2}{a^6}, \quad V_0 = {\rm Const.} \label{Kinva}
\end{equation}
The relation \eqref{Kinva} holds for $K = I$ both for massive and
massless spinor, whereas, in case of $K$ being one of the
expressions $\{J,\, I + J,\, I - J\}$ it is true for the massless
spinor field only.

From \eqref{SF1frw} for the invariants of bilinear spinor form in
this case we have
\begin{subequations}
\label{invfrw}
\begin{eqnarray}
\dot S_0  +  {\cG} A_{0}^{0} &=& 0, \label{S0frw} \\
\dot P_0  -  \Phi A_{0}^{0} &=& 0, \label{P0frw}\\
\dot A_{0}^{0}  +  \Phi P_0 -  {\cG}
S_0 &=& 0, \label{A00frw}\\
\dot A_{0}^{3}  &=& 0, \label{A03frw}\\
\dot v_{0}^{0}  &=& 0,\label{v00frw} \\
\dot v_{0}^{3} +
\Phi Q_{0}^{30} +  {\cG} Q_{0}^{21} &=& 0,\label{v03frw}\\
\dot Q_{0}^{30} -  \Phi v_{0}^{3} &=& 0,\label{Q030frw} \\
\dot Q_{0}^{21} -  {\cG} v_{0}^{3} &=& 0. \label{Q021frw}
\end{eqnarray}
\end{subequations}
where we denote $S_0 = S a^3,\, P_0 = P a^3,\, A_0^\mu = A_\mu
a^3,\, v_0^\mu = v^\mu a^3,\, Q_0^{\mu \nu} = Q^{\mu \nu} a^3$. Here
$v^\mu = \bp \gamma^\mu \psi$, \,\,\,\,  $A^\mu = \bp \gamma^5
\gamma^\mu \psi$ and $ Q^{\mu\nu} = \bp \sigma^{\mu\nu} \psi$  are
the vector, pseudovector and tensor bilinear spinor forms,
respectively.

From (\ref{S0frw}) - (\ref{Q021frw}) one finds the following
relations:
\begin{subequations}
\label{inv0frw}
\begin{eqnarray}
(S_{0})^{2} + (P_{0})^{2} + (A_{0}^{0})^{2}  &=&
C_1 = {\rm Const}, \label{inv01frw}\\
A_{0}^{3} &=& C_2 = {\rm Const}, \label{inv02frw}\\
v_{0}^{0} &=& C_3 = {\rm Const}, \label{inv03frw}\\
(Q_{0}^{30})^{2} + (Q_{0}^{21})^{2} + (v_{0}^{3})^{2}
 &=& C_4 = {\rm Const}. \label{inv04frw}
\end{eqnarray}
\end{subequations}

It should be noted that in case of a Bianchi type-I space-time for
bilinear spinor forms one obtains the system that is exactly the
same as \eqref{invfrw} with the solution \eqref{inv0frw}. But the
presence of non-trivial non-diagonal components of energy-momentum
tensor of the spinor field in that case leads to $A^1 = A^2 = A^3 =
0 $ if the restrictions are imposed on the spinor fields only. And
in that specific case thanks to Fierz identity the relation $A_\mu
v^\mu = 0$ leads to $S = 0$ and $P = 0$, hence vanishing mass and
nonlinear terms in the spinor field Lagrangian \cite{Saha2015APSS}.
But the absence of non-trivial non-diagonal components of the
energy-momentum tensor of spinor field in FRW model does not lead to
such a severe situation.

\section{Solution to the field equations}

In this section we will solve the field equations derived in the
previous section.

\subsection{Solution to the spinor field equations}

Let us now solve the spinor field equation. We consider the
4-component spinor field given by
\begin{eqnarray}
\psi = \left(\begin{array}{c} \psi_1\\ \psi_2\\ \psi_3 \\
\psi_4\end{array}\right). \label{psi}
\end{eqnarray}

Denoting $\phi_i = a^{3/2} \psi_i$ the spinor field equation
\eqref{speq1pfrw} can be written as

\begin{subequations}
\label{speq1pfrwpsi}
\begin{eqnarray}
\dot \phi_1 + \imath\,\left(m_{\rm sp} + \cD\right) \phi_1 +  {\cG} \phi_3 &=& 0, \label{ph1frw}\\
\dot \phi_2 + \imath\, \left(m_{\rm sp} + \cD\right) \phi_2 +  {\cG} \phi_4 &=& 0, \label{ph2frw}\\
\dot \phi_3 - \imath\, \left(m_{\rm sp} + \cD\right) \phi_3 -  {\cG} \phi_1 &=& 0, \label{ph3frw}\\
\dot \phi_4 - \imath\, \left(m_{\rm sp} + \cD\right) \phi_4 -  {\cG}
\phi_2&=& 0, \label{ph4frw}
\end{eqnarray}
\end{subequations}

We study the cases for different $K$ in details.

In case of $K = I$ for the components of spinor field from
\eqref{speq1pfrwpsi} one finds \cite{Saha2001PRD}
\begin{subequations}
\label{psinlfrw}
\begin{eqnarray}
\psi_1(t) &=& \frac{C_1}{{a^{3/2}}} \exp \,\left(-\imath\, \left[m_{\rm sp} + {\cD}\right] dt\right),\\
\psi_2(t) &=& \frac{C_2}{{a^{3/2}}} \exp\,\left(-\imath\, \left[m_{\rm sp} + {\cD}\right] dt\right),\\
\psi_3(t) &=& \frac{C_3}{{a^{3/2}}} \exp\,\left(\imath\, \left[m_{\rm sp} + {\cD}\right] dt\right),\\
\psi_4(t) &=& \frac{C_4}{{a^{3/2}}} \exp\,\left(\imath\,
\left[m_{\rm sp} + {\cD}\right] dt\right),
\end{eqnarray}
\end{subequations}
with $C_1,\,C_2,\,C_3,\,C_4$ being the integration constants and
related to $V_0$ as
$$C_1^* C_1 + C_2^* C_2 - C_3^* C_3 - C_4^* C_4 = V_0,$$
where $V_0$ such that $I = V_0^2/a^6.$

As it was mentioned earlier for $K = J$ only a massless spinor field
provides exact solution \cite{Saha1997GRG}. For the massless spinor
field in this case we find \cite{Saha2001PRD}
\begin{subequations}
\label{psij}
\begin{eqnarray}
\psi_1 &=&\frac{1}{{a^{3/2}}} \left(D_1 e^{i \sigma} +
iD_3 e^{-i\sigma}\right), \\
\psi_2 &=&\frac{1}{{a^{3/2}}} \left(D_2 e^{i \sigma} +
iD_4 e^{-i\sigma}\right),  \\
\psi_3 &=&\frac{1}{{a^{3/2}}} \left(iD_1 e^{i \sigma} +
D_3 e^{-i \sigma}\right), \\
\psi_4 &=&\frac{1}{{a^{3/2}}} \left(iD_2 e^{i \sigma} + D_4
e^{-i\sigma}\right),
\end{eqnarray}
\end{subequations}
with the constants $D_i$s obeying $D_1^* D_1 + D_2^* D_2 - D_3^* D_3
- D_4^* D_4 = V_0,$ where $J = V_0^2/a^6$.

As far as cases with $K = I + J$ or $K = I - J$ are concerned, the
solution to the equation \eqref{speq1pfrwpsi} can be written in the
form \cite{Saha2016EPJP}
\begin{equation}
\phi(t) =  T\, \exp{\left(-\int_t^{t_1}  A (\tau) d \tau\right) \phi
(t_1)}, \label{phiIpmJ}
\end{equation}
where
\begin{equation}
A = \left(\begin{array}{cccc}-\imath\, {\cD} &0 & - {\cG}& 0
\\ 0&-\imath\, {\cD}& 0 &- {\cG}\\ {\cG}&0&\imath\, {\cD}&0\\0& {\cG} & 0
&\imath\, {\cD} \end{array}\right). \label{AMat1frw}
\end{equation}
and $\phi (t_1) = {\rm
col}\left(\phi_{1}^{0},\,\phi_{2}^{0},\,\phi_{3}^{0},\,\phi_{4}^{0}\right)$
is the solution at $t = t_1$ with $\phi_i^0$ being some constants.

Thus the spinor field equations are totally solved and it was found
that the components of spinor field are inverse to $a^{3/2}$.

\subsection{Solutions to the gravitational field equations}

Let us now go back to the equation \eqref{frw0}. We choose the
polynomial nonlinearity in the from \cite{Saha2015EPJP}
\begin{equation}
F = \sum_{k} \lambda_k K^{n_k}, \label{nonlinearity}
\end{equation}
where $\lambda_k$ are the coupling constants. For simplicity we
consider the case with $K = I$. Taking into account that $S =
V_0/a^3$ ($I = V_0^2/a^6$) we now have

\begin{equation}
\ddot a = \frac{\kappa}{6}\left(2 \sum_{k} \lambda_k \left(1 - 3
n_k\right)\frac{V_0^{2 n_k}}{a^{6n_k}}  - m_{\rm sp} \frac{V_0}{a^3}
\right)\, a. \label{frw01}
\end{equation}

We solve this equation numerically. Since at any space-time point
where $a = 0$ the invariants of spinor and gravitational fields
become singular, it is a singular point \cite{Saha2001PRD}. So, we
consider the initial value of $a (0)$ is small but non-zero. As a
result for the nonlinear term to prevail at the early stage of
evolution, in \eqref{frw01} we should set $ n_k = n_1: 1 - 6 n_1 <
-2$, i.e., $n_1
> 1/2$, whereas for an expanding Universe when $a \to \infty$ as $t
\to \infty $ one should have $ n_k = n_2: 1 - 6 n_2 > -2$, i.e.,
$n_2 < 1/2$. On account of this we can write the foregoing equation
as
\begin{eqnarray}
\ddot a &=& \Phi(a), \label{aPhi}\\
\Phi(a)&=& \frac{\kappa}{6}\left(2 \lambda_1 \left(1 - 3
n_1\right)\frac{V_0^{2 n_1}}{a^{6n_1}}  +  2 \lambda_2 \left(1 - 3
n_2\right)\frac{V_0^{2 n_2}}{a^{6n_2}} - m_{\rm sp} \frac{V_0}{a^3}
\right)\, a, \nonumber
\end{eqnarray}
with the first integral
\begin{eqnarray}
\dot a &=& \Phi_1(a), \label{aPhi1}\\
\Phi_1(a)&=& \sqrt{\frac{\kappa}{3}\left(V_0^{2 n_1} \lambda_1 a^{2(
1 - 3n_1)}  +  V_0^{2 n_2} \lambda_2 a^{2( 1 - 3n_2)} + m_{\rm sp}
\frac{V_0}{a} \right)+ \bar C}, \nonumber
\end{eqnarray}
with $\bar C$ being an integration constant.

\begin{figure}[ht]
\centering
\includegraphics[height=70mm]{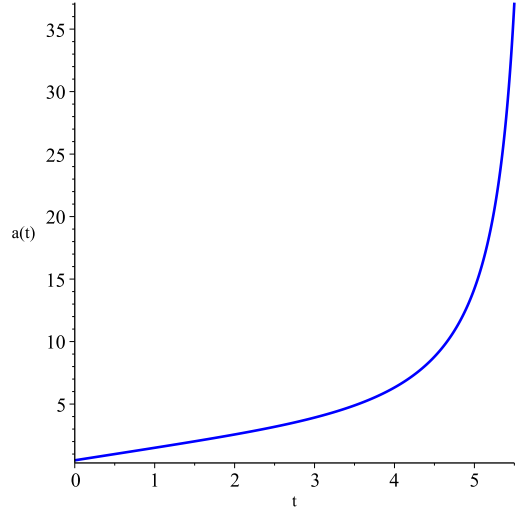} \\ \vskip 1 cm
\caption{Evolution of the Universe for a positive  $\lambda_2$}
\label{FRWVl1pl2p}
\end{figure}

\begin{figure}[ht]
\centering
\includegraphics[height=70mm]{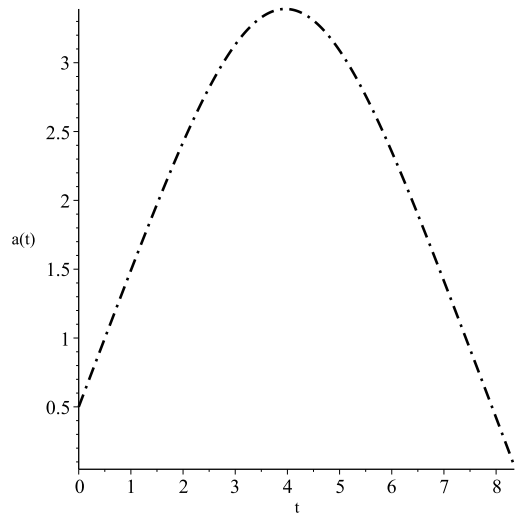} \\ \vskip 1 cm
\caption{Evolution of the Universe for a negative  $\lambda_2$}
\label{FRWVl1pl2m}
\end{figure}

\begin{figure}[ht]
\centering
\includegraphics[height=70mm]{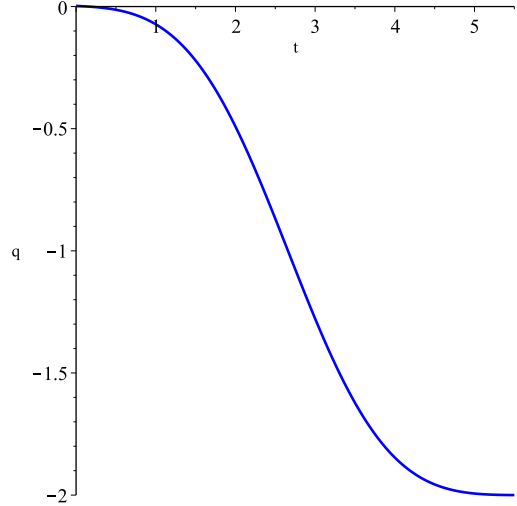} \\ \vskip 1 cm
\caption{Plot of deceleration parameter $q$ for a positive
$\lambda_2$ } \label{FRWql1pl2p}
\end{figure}

\begin{figure}[ht]
\centering
\includegraphics[height=70mm]{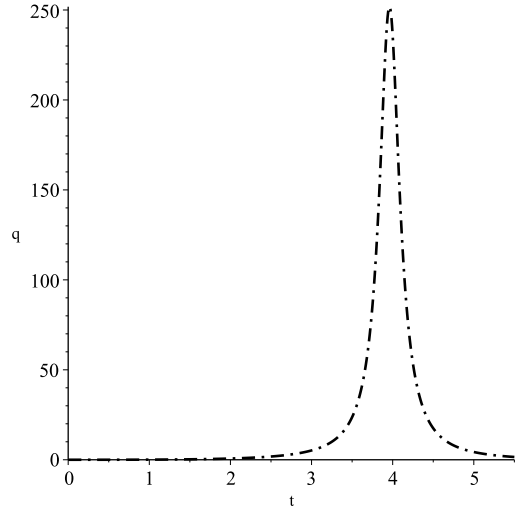} \\ \vskip 1 cm
\caption{Plot of deceleration parameter $q$ for a negative
$\lambda_2$ } \label{FRWql1pl2m}
\end{figure}

In what follows we study the equation \eqref{aPhi} numerically. As
it was mentioned earlier, we set a small but non-zero initial value
for $a(t)$, precisely, $a(0) = 0.5$, while $\dot a(t)$ is calculated
from \eqref{aPhi1}. We also set $V_0 = 1$, $m_{\rm sp} = 0.01$,
$\bar C = 1$, $\kappa = 1$, $n_1 = 2/3$, and  $n_2 = -1/6$. We have
considered both positive and negative values for $\lambda_1$
($\lambda_1 = \pm 0.0001$) and $\lambda_2$ ($\lambda_2 = \pm 0.03$).
It is found that the sign of $\lambda_1$ has almost no effect, while
that of $\lambda_2$ is crucial. A positive $\lambda_2$ generates an
expanding Universe [cf. Fig. \ref{FRWVl1pl2p}], while a negative
$\lambda_2$ gives rise to an Universe that after attaining some
maximum begins to contract and finally ends in Big Crunch [cf. Fig.
\ref{FRWVl1pl2m}]. In Figs. \ref{FRWql1pl2p} and \ref{FRWql1pl2m}
the corresponding pictures of evolution of deceleration parameter
are given.

\section{Spinor description of matter}

One of the principal advantage of using spinor description of source
field lies on the fact that in this case the Bianchi identity is
fulfilled automatically. To solve a self-consistent system beside
Einstein field equations one has to consider the Bianchi identity
that gives relation between material field and geometry. The spinor
description of source field gives a straightforward exit when a
multiple sources are considered. To show that let us write the
Biacnhi identity $G_{\mu;\nu}^{\nu} = 0$, that leads to

\begin{equation}
T_{\mu;\nu}^{\nu} = T_{\mu,\nu}^{\nu} + \G_{\rho \nu}^{\nu}
T_\mu^\rho -   \G_{\mu \nu}^{\rho} T_\rho^\nu = 0,
\label{frwidentity}
\end{equation}
which for the metric \eqref{frw} on account of the components of the
energy-momentum tensor takes the from
\begin{equation}
\dot \varepsilon + 3 \frac{\dot a}{a}\left(\varepsilon + p\right) =
0. \label{Conserv}
\end{equation}
Inserting $\varepsilon$ and $p$ from \eqref{emt00frw} and
\eqref{emtiifrw} from \eqref{Conserv} one finds

\begin{equation}
\frac{m_{\rm sp}}{S}\,\, \frac{d}{dt}\left( S a^3 \right) +
\frac{F_K}{a^6} \,\, \frac{d}{dt}\left( K a^6 \right) = 0.
\label{Conserv1}
\end{equation}

In case of $K = I = S^2$ \eqref{Conserv1} fulfills identically as $S
a^3 = const.$ and $K a^6 = const.$, whereas in the case when $K$
takes one of the following expressions $\{J,\,I+J,\,I-J\}$, for a
massless spinor field \eqref{Conserv1} fulfills identically as $K
a^6 = const.$ Hence if we use spinor description of different fluid
and dark energy simulated from corresponding equation of sate, the
Bianchi identity will be fulfilled identically without invoking any
additional condition.

Let us now consider the case with massless spinor field when it can
successfully simulate different types of fluid and dark energy. Here
we simply give the expressions corresponding to different matters. A
detailed description of how to obtain these expressions can be found
in \cite{Saha2014IJTP,Saha2015APSS}.

\subsection{perfect fluid and quintessence}

Let us first consider the case when the spinor filed generates a
fluid with barotropic equation of state

\begin{equation}
p = W \ve, \quad W = {\rm const.} \label{beos}
\end{equation}

\begin{figure}[ht]
\centering
\includegraphics[height=70mm]{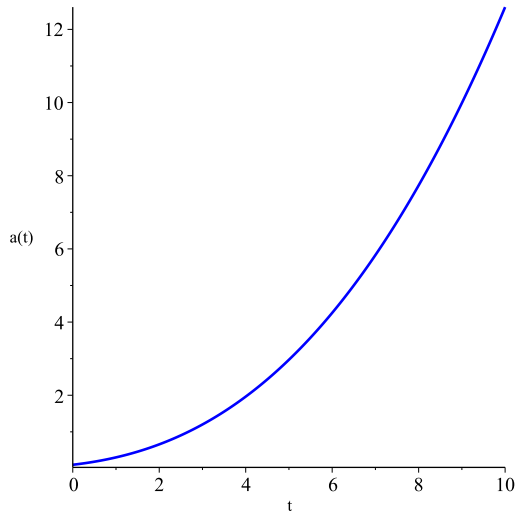} \\ \vskip 1 cm
\caption{Plot of metric function $a$ in case of a quintessence.}
\label{FRWq}
\end{figure}

In this case from \eqref{Conserv} we have

\begin{equation}
F(S) = \lambda S^{1+W}, \quad \lambda = {\rm const.} \label{sol1}
\end{equation}
Depending on the value of $W$ the nonlinearity \eqref{sol1} gives
rise to dust ($W = 0$), radiation ($W = 1/3$), hard Universe ($W \in
(1/3,\,1)$), stiff matter ($W = 1$), quintessence ($W \in
(-1/3,\,-1)$), cosmological constant ($W = -1$), phantom matter ($W
< -1$) and ekpyrotic matter ($W > 1$). In Fig. \ref{FRWq} we have
given the evolution of $a$ when the Universe is filled with
quintessence setting $W = -1/2$. It should be mentioned that if one
considers a massive spinor field, the nonlinear term obtained from
\eqref{Conserv} in that case contains a term that cancels the mass
term in the spinor field Lagrangian
\cite{Saha2014IJTP,Saha2015APSS}.

\subsection{Chaplygin gas}

\begin{figure}[ht]
\centering
\includegraphics[height=70mm]{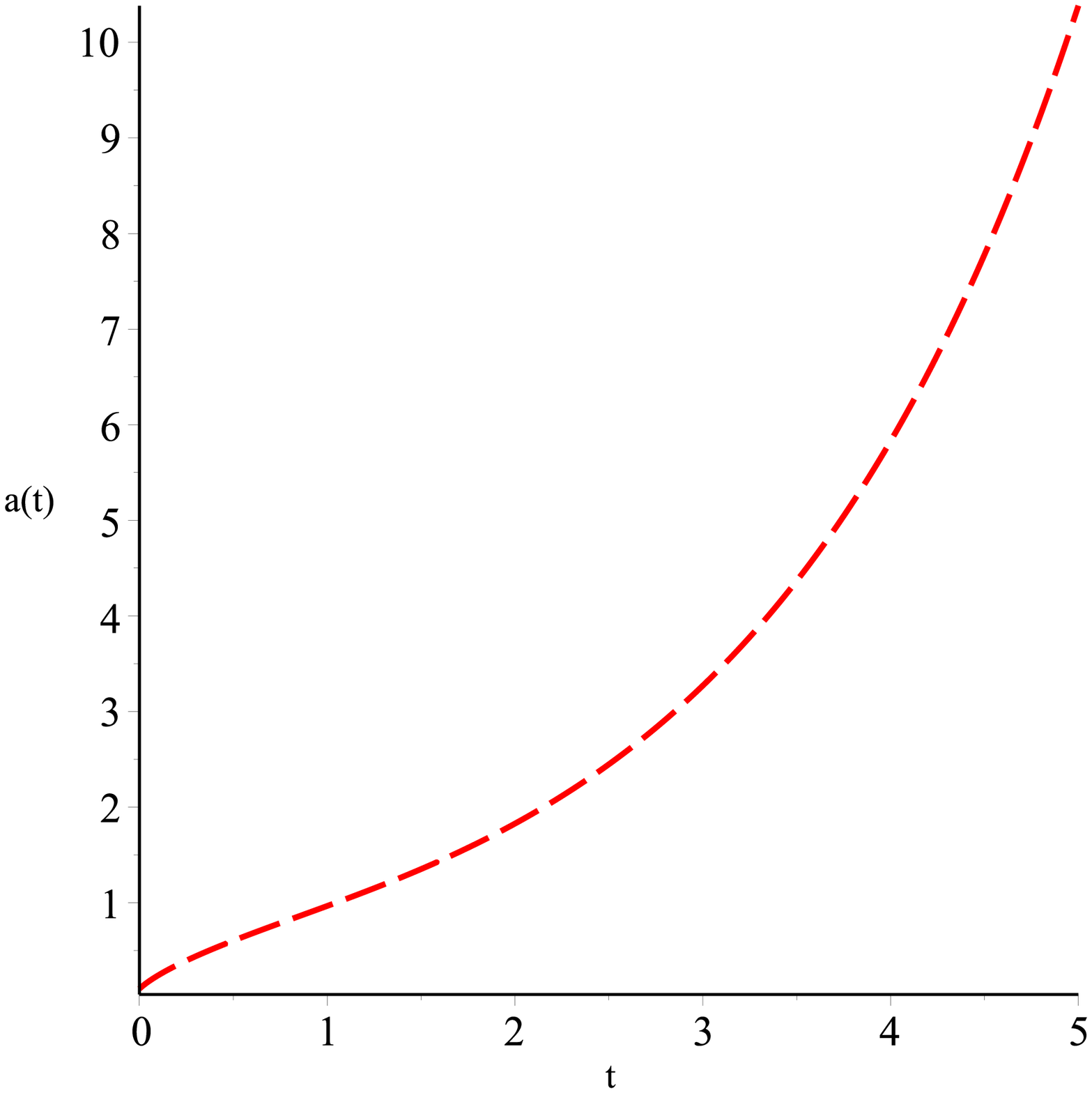} \\ \vskip 1 cm
\caption{Plot of metric function $a$ in case of a Chaplygin gas.}
\label{FRWch}
\end{figure}

A Chaplygin gas can be given by
\begin{equation}
p_{\rm ch} = - \frac{A}{\ve_{\rm ch}}, \label{chap}
\end{equation}
which corresponds to the nonlinearity
\begin{equation}
F = \left(A + \lambda K^{(1+\alpha)/2}\right)^{1/(1+\alpha)}.
\label{chapsp}
\end{equation}
with $ A $ being a positive constant and $ 0 <\alpha \le 1. $ In
this case evolution of $a$ is given in Fig. \ref{FRWch}

\subsection{Oscillating dark energy}

\begin{figure}[ht]
\centering
\includegraphics[height=70mm]{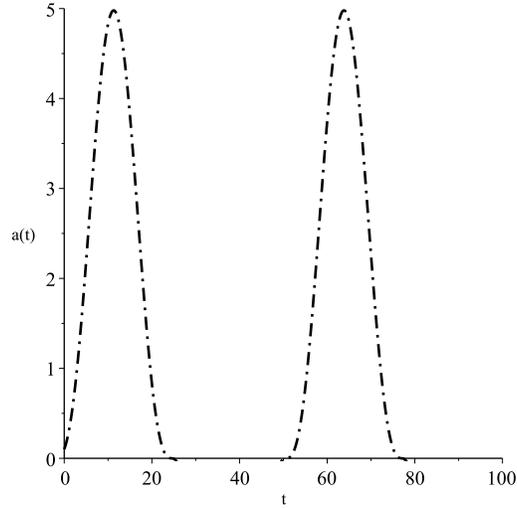} \\ \vskip 1 cm
\caption{Plot of metric function $a$ in case of a modified
quintessence.} \label{FRWmq}
\end{figure}

An oscillating dark energy (modified quintessence) with EoS
\cite{Saha2006IJTP}
\begin{eqnarray}
p =  W (\ve - \ve_{\rm cr}), \quad W \in (-1,\,0), \label{mq}
\end{eqnarray}
can be simulated by
\begin{equation}
F = \lambda K^{(1+W)/2} + \frac{W}{1+W}\ve_{\rm cr}, \label{Fmq}
\end{equation}
with $\ve_{\rm cr}$ being some constant. In this case we have a
oscillatory mode of expansion given in Fig. \ref{FRWmq}

\begin{figure}[ht]
\centering
\includegraphics[height=70mm]{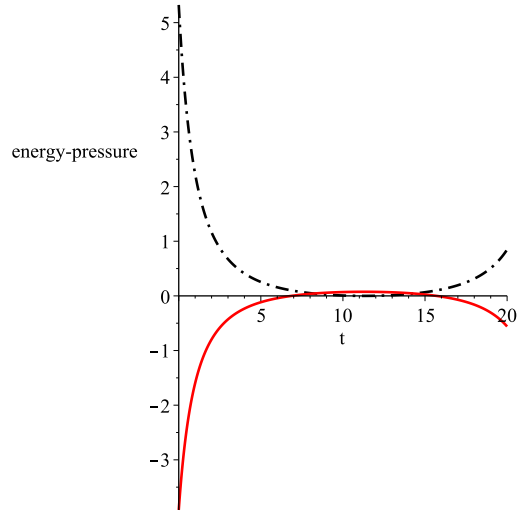} \\ \vskip 1 cm
\caption{Plot of energy density (black, dash-dot line) and pressure
(red, solid line) in case of a modified quintessence.}
\label{FRWmqen-pr}
\end{figure}

Corresponding energy density and pressure is given in Fig.
\ref{FRWmqen-pr}. As one sees, as the energy density becomes less
than $\ve_{\rm cr}$, pressure becomes positive. It results in
contraction of space-time, hence the increase of energy density.
This leads to pressure becoming negative and hence the Universe
expanding very fast. Thus we see that in this case the Universe
begins from initial singularity (Big Bang), expands to some maximum,
then begin to contract and ends in Big Crunch only to begin from a
Big Bang.

\subsection{Modified Chaplygin gas}

In case of a modified Chaplygin gas with equation of state

\begin{eqnarray}
p = W \ve - A/\ve^\alpha, \label{mchap}
\end{eqnarray}

we have

\begin{eqnarray}
F = \left(\frac{A}{1+W} + \lambda
S^{(1+\alpha)(1+W)}\right)^{1/(1+\alpha)}. \label{mchapsp}
\end{eqnarray}

\begin{figure}[ht]
\centering
\includegraphics[height=70mm]{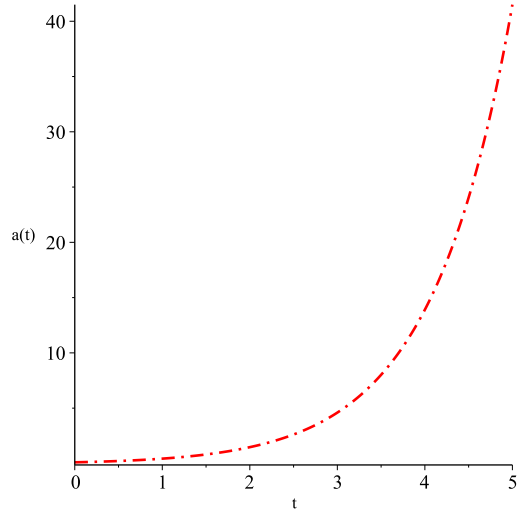} \\ \vskip 1 cm
\caption{Plot of metric function $a$ in case of a modified Chaplygin
gas.} \label{FRWmch}
\end{figure}

The corresponding evolution is given in Fig. \ref{FRWmch}

\subsection{Quintom}

In order to understand the behavior of the equation of state of dark
energy \eqref{beos} with $ w_{\rm quint}> -1 $ in the past and with
$ w_{\rm quint} < -1 $ at present, quintom model of dark energy was
proposed \cite{Feng2005}. Quintom model is a dynamic model of dark
energy and compare to the other models of dark energy it defines the
cosmic evolution in a different way. One of the characteristics of
Quintom model is the fact that its equation of state can smoothly
pass the value of $ w_{\rm quint} = -1 $ \cite{Cai2007}. In contrast
to \eqref{beos}, where $w$ is a constant, in Quintom model it
depends on time and can be given by the EoS
\begin {equation}
w_{\rm quint}(t) = - r - \frac{s}{t^2}, \label{quintom}
\end{equation}
where $r$ and $s$ are some parameters. Many authors have used
Quintom model in order to generate a bouncing Universe. Spinor
description of Quintom model was given in \cite{Cai2008}. Following
\cite{Cai2008} let us construct corresponding spinor field
nonlinearity. In doing so let us first write the EoS parameter.
Taking into account that $\ve = T_0^0$ and $p = - T_1^1$ from
\eqref{emt00frw} and \eqref{emtiifrw} for the massless spinor field
we have
\begin {equation}
w_{\rm quint} = \frac{p}{\ve} = \frac{2 K F_K - F(K)}{F(K)} = - 1 +
\frac{2 K F_K}{F}. \label{quintom1}
\end{equation}
Since energy density $\ve$ should be positive, $F(K) = \ve$ should
be positive as well. For $w_{\rm quint} > -1$ then we should have
$F_K > 0$, while for  $w_{\rm quint} < -1$ we should have $F_K < 0$.
Moreover from \eqref{Kinva} we see that for an expanding Universe
$K$ is a decreasing function of time. Taking all this into account
we like in \cite{Cai2008} construct 3 scenario:

\vskip 5 mm {\bf (1)} quintom-A scenario that describes the Universe
evolving from quintessence phase with $w_{\rm quint} > -1$ to a
phantom phase with $w_{\rm quint} < -1$. In this case we have
$$ F_K  > 0 \quad  \to  \quad F_K < 0;$$

As a study case we can consider

\begin{equation}
F (K) = \lambda\left[(K - b)^2 + c\right], \quad F_K = 2  \lambda( K
- b), \label{quintom1a0}
\end{equation}
where $b$ and $c$ are some positive constants. In this case $F (K)
> 0$. As far as $F_K$ is concerned, at the initial
stage when the Universe is small enough, i.e. $V << 1$, $ K =
V_0^2/V^2$ is quite large, hence $F_K > 0$. This leads to $w_{\rm
quint} > -1$, i.e. we have quintessence like phase. At $K = b$ we
have $F_K = 0$, with EoS $w_{\rm quint} = -1$.  After that as the
Universe expands $K$ becomes small, hence $F_K < 0$. As a result we
have phantom-like phase with $w_{\rm quint} < -1$.  Under this
choice we have

\begin {equation}
w_{\rm quint} = -1 + \frac{4 K(K - b)}{\left[(K - b)^2 + c\right]}.
\label{quintom1a}
\end{equation}

\begin{figure}[ht]
\centering
\includegraphics[height=70mm]{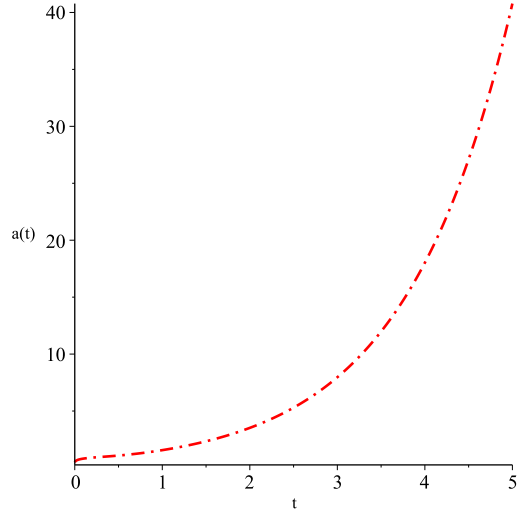} \\ \vskip 1 cm
\caption{Plot of metric function $a$ in case of a type 1 quintom
scenario.} \label{FRWquintom1}
\end{figure}

\begin{figure}[ht]
\centering
\includegraphics[height=70mm]{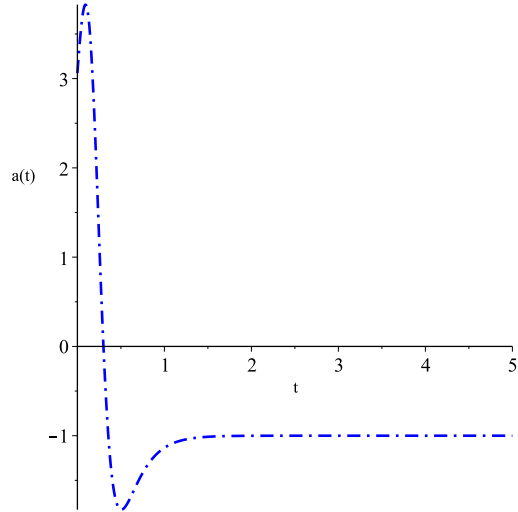} \\ \vskip 1 cm
\caption{Plot of EoS parameter in case of a type 1 quintom
scenario.} \label{FRWquintom1eos}
\end{figure}

Corresponding behavior of metric function $a$ and EoS parameter are
given in Figs. \ref{FRWquintom1} and \ref{FRWquintom1eos}

\vskip 5 mm {\bf (2)}  quintom-B scenario that describes the
Universe evolving from phantom phase with $w_{\rm quint} < -1$ to a
quintessence phase with $w_{\rm quint} > -1$. In this case we have

$$ F_K  < 0 \quad  \to  \quad F_K > 0;$$

In this case Cai and Wang \cite{Cai2008} proposed $F = \lambda
\left[-(K - b)K + c\right]$. Though in this case the condition $ F_K
< 0 \quad  \to  \quad F_K > 0$ fulfills, the function $F$ hence the
energy density becomes negative at the early stage of evolution. So
we propose
\begin{equation}
F (K) = \frac{\lambda}{\left[(K - b)^2 + c\right]}, \quad F_K =
-\frac{2 \lambda(K - b)}{\left[(K - b)^2 + c\right]^2}.
\label{quintom1b0}
\end{equation}
As one sees, $ F (K)$ is always positive. From \eqref{quintom1b0} it
can be easily verified that at the initial stage when $K$ in large,
$F_K < 0$. It is a phantom-like phase with $w_{\rm quint} < -1$.
With the expansion of the Universe $K$ becomes small. At $K = b$ we
have $F_K = 0$, that is  $w_{\rm quint} = -1$. Further as $K$
decreases, $F_K$ becomes positive, thus giving rise to
quintessence-like phase with $w_{\rm quint}
> -1$. In this case we have

\begin {equation}
w_{\rm quint} = - 1 - \frac{4 K (K- b)}{\left[(K - b)^2 + c\right]}.
\label{quintom1b}
\end{equation}

\begin{figure}[ht]
\centering
\includegraphics[height=70mm]{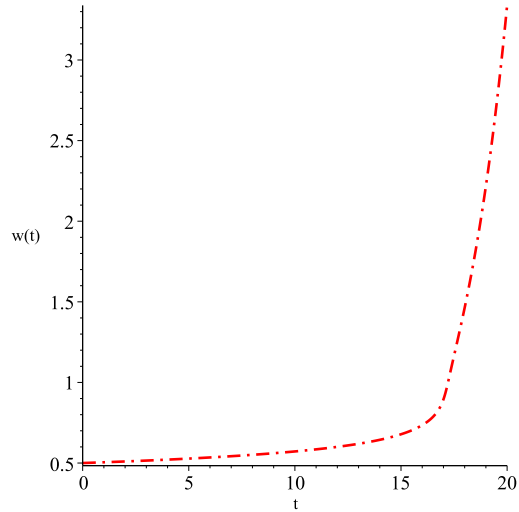} \\ \vskip 1 cm
\caption{Plot of metric function $a$ in case of a type 2 quintom
scenario.} \label{FRWquintom2}
\end{figure}

\begin{figure}[ht]
\centering
\includegraphics[height=70mm]{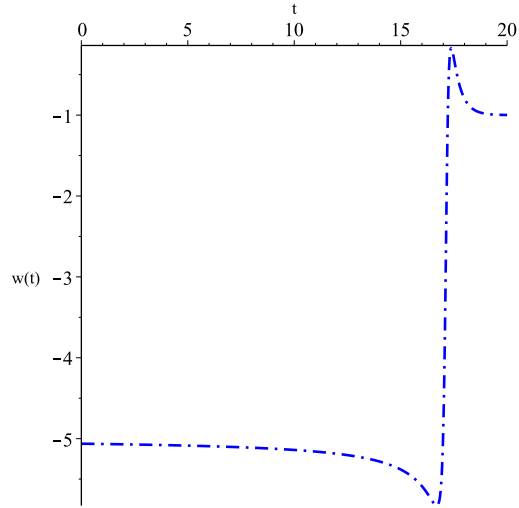} \\ \vskip 1 cm
\caption{Plot of  EoS parameter in case of a type 2 quintom
scenario.} \label{FRWquintom2eos}
\end{figure}

Corresponding behavior of metric function $a$ and EoS parameter are
given in Figs. \ref{FRWquintom2} and \ref{FRWquintom2eos}

\vskip 5 mm {\bf (3)} quintom-C scenario when $F_K$ changes its sign
more than one time. In this case one can obtain a new quintom
scenario with EoS crossing $-1$ many times. In this case one can set

\begin{equation}
F (K) = \lambda\left[ K(K - b)^2 + c\right], \quad F_K = \lambda(K -
b)(3K - b). \label{quintom1c0}
\end{equation}

From \eqref{quintom1c0} we again have $F(K) > 0$ and $F_K$ changes
its sign more than once. In this case

\begin {equation}
w_{\rm quint} = - 1 - \frac{2 K (K- b)(3K - b)}{\left[K(K - b)^2 +
c\right]}. \label{quintom1c}
\end{equation}

Note that this case can be simulated with $F(K) = c + \sin (K)$,
where $c > 1$, hence $F(K) > 0$. In this case $F_K = \cos(K)$ is the
alternating quantity.

\begin{figure}[ht]
\centering
\includegraphics[height=70mm]{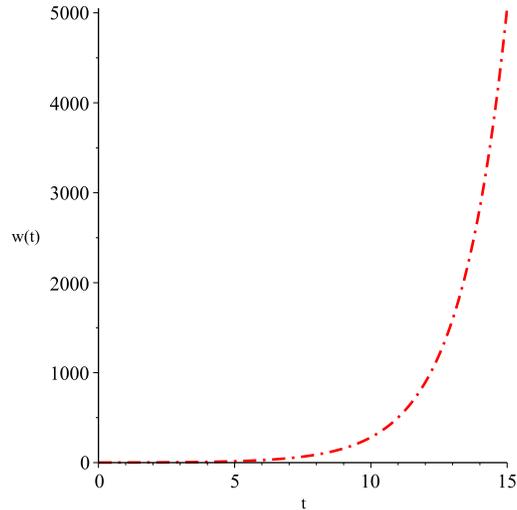} \\ \vskip 1 cm
\caption{Plot of metric function $a$ in case of a type 3 quintom
scenario.} \label{FRWquintom3}
\end{figure}

\begin{figure}[ht]
\centering
\includegraphics[height=70mm]{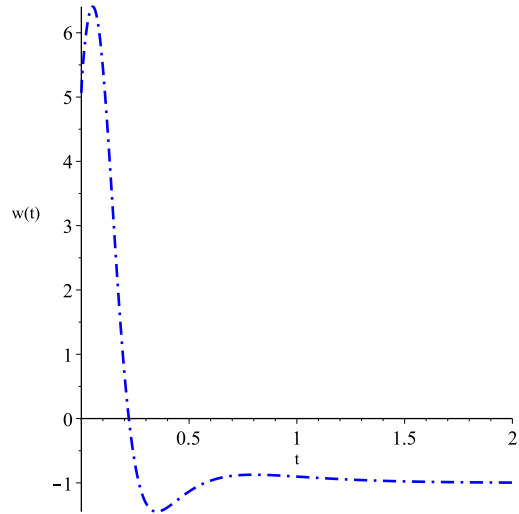} \\ \vskip 1 cm
\caption{Plot of  EoS parameter in case of a type 3 quintom
scenario.} \label{FRWquintom3eos}
\end{figure}

Corresponding behavior of metric function $a$ and EoS parameter are
given in Figs. \ref{FRWquintom3} and \ref{FRWquintom3eos}

\subsection{Why spinor description is convenient?}

Question may arise "Why does one need spinor description of matter?"
To answer this question, let us recall that the Universe was filled
with different matter in the course of evolution. While in the past
the evolution was matter dominated, now the dominant component is
the dark energy. In case when there is only two components, say
radiation and dark energy, the continuity equation can be easily
separated thanks to the fact that the dark energy does not interact
with usual matter. But in case of three or more components the
situation becomes complicated. Exploiting the spinor description one
may avoid this situation, as the continuity equation for spinor
field is fulfilled identically [cf. \eqref{Conserv1} and discussion
thereafter]. To demonstrate it, let us consider the case with a
Van-der-Waals gas, radiation and modified Chaplygin gas. In this
case we have
\begin{eqnarray}
\ve &=& \ve_{\rm vdw} + \ve_{\rm rad} + \ve_{\rm mchap}, \label{3in1}\\
p &=& p_{\rm vdw} + p_{\rm rad}  + p_{\rm mchap} = \frac{8 W_{\rm
vdw} \ve_{\rm vdw}}{3 - \ve_{\rm vdw}} - 3 \ve_{\rm vdw}^2 + W_{\rm
rad} \ve_{\rm rad} + W_{\rm mchap} \ve_{\rm mchap} - A/ \ve_{\rm
mchap}^\alpha. \nonumber
\end{eqnarray}
Acting straightforward, from \eqref{Conserv} in this case we obtain
\begin{equation}
\dot \ve_{\rm vdw} + \dot \ve_{\rm rad} + \dot  \ve_{\rm mchap} + 3
\frac{\dot a}{a} \left[\left(\ve_{\rm vdw} + p_{\rm vdw}\right) +
\left(\ve_{\rm rad} + p_{\rm rad}\right) + \left(\ve_{\rm mchap} +
p_{\rm mchap} \right) \right] = 0 \label{3in1comp}
\end{equation}
Now assuming that the dark energy does not interact with normal
matter, the foregoing equation can be decomposed to the following
system:
\begin{subequations}
\begin{eqnarray}
\dot \ve_{\rm vdw} + \dot \ve_{\rm rad} + 3 \frac{\dot a}{a}
\left[\left(\ve_{\rm vdw} + p_{\rm vdw}\right) + \left(\ve_{\rm rad}
+ p_{\rm rad}\right)
\right] &=& 0, \label{3in1comp1}\\
 \dot  \ve_{\rm mchap} + 3 \frac{\dot a}{a} \left(\ve_{\rm mchap} + p_{\rm mchap}
 \right) &=& 0. \label{3in1comp2}
\end{eqnarray}
\end{subequations}
Thus we have only two equations for three unknowns, i.e., the system
is underdetermined. Whereas, expressing only two of the three
components (say, radiation and modified Chaplygin gas) in terms of
spinor field nonlinearity we can easily overcome this situation.

\begin{figure}[ht]
\centering
\includegraphics[height=70mm]{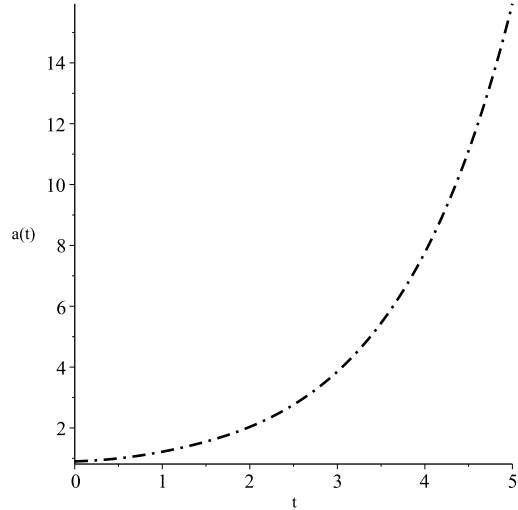} \\ \vskip 1 cm
\caption{Plot of metric function $a$ in case of the FRW Universe
filled with Van-der-Waals gas, radiation and modified Chaplygin
gas.} \label{a3n1}
\end{figure}

\begin{figure}[ht]
\centering
\includegraphics[height=70mm]{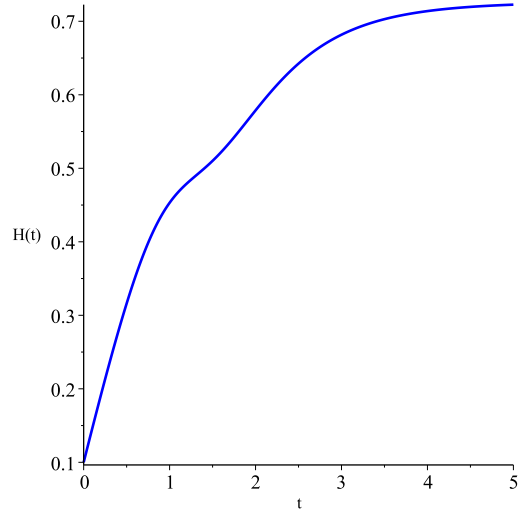} \\ \vskip 1 cm
\caption{Plot of Hubble parameter in case of the FRW Universe filled
with Van-der-Waals gas, radiation and modified Chaplygin gas.}
\label{Hubble3n1}
\end{figure}

\begin{figure}[ht]
\centering
\includegraphics[height=70mm]{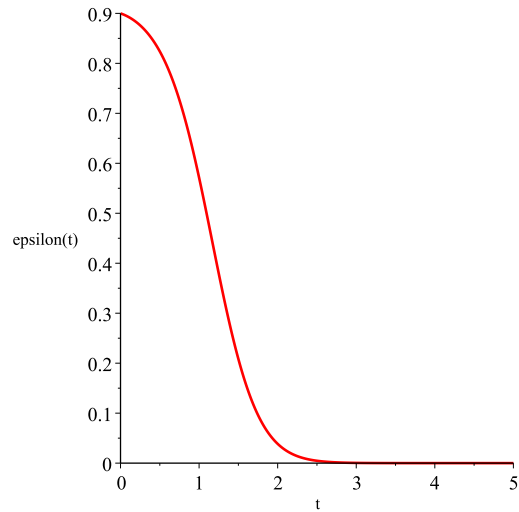} \\ \vskip 1 cm
\caption{Evolution of energy density of a Van-der-Waals gas}
\label{endenvdw}
\end{figure}

\begin{figure}[ht]
\centering
\includegraphics[height=70mm]{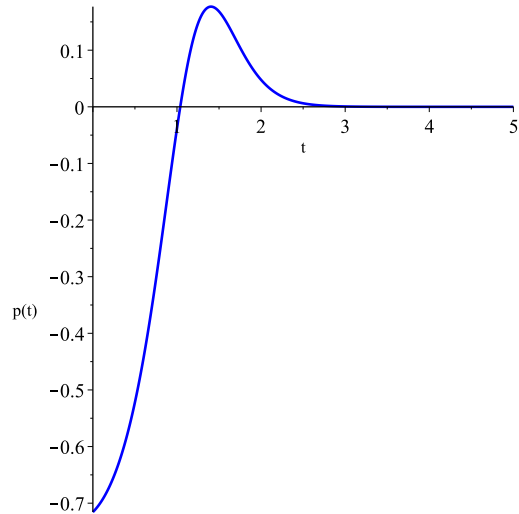} \\ \vskip 1 cm
\caption{Evolution of pressure of a Van-der-Waals gas}
\label{pressvdw}
\end{figure}

\begin{figure}[ht]
\centering
\includegraphics[height=70mm]{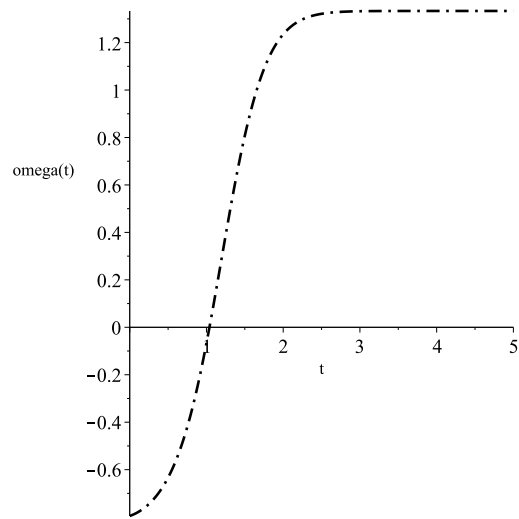} \\ \vskip 1 cm
\caption{Evolution of EoS parameter of a Van-der-Waals Gas}
\label{eosvdw}
\end{figure}

Further taking into account that radiation and modified Chaplygin
gas we can simulate by means of spinor field nonlinearity as
\begin{eqnarray}
\ve_{\rm rad} = \lambda_{\rm rad} S^{4/3}, \quad \ve_{\rm mchad} =
\left( \frac{A}{1 + W}+\lambda_{\rm mchap}
S^{(1+W)(1+\alpha)}\right)^{1/(1+\alpha)}. \label{radmchap}
\end{eqnarray}

the equation \eqref{frw0} together with the continuity equations can
be written as

\begin{subequations}
\begin{eqnarray}
\dot H &=& - H^2 - \frac{\kappa}{6} f(a,\, \ve_{\rm vdw}), \label{sysH}\\
\dot a &=& a H, \label{sysa}\\
\dot \ve_{\rm vdw} &=& - 3 H \left(\ve_{\rm vdw} + \frac{8 W_{\rm
vdw}\ve_{\rm vdw}}{3 - \ve_{\rm vdw}} - 3 \ve_{\rm vdw}^2\right),
\label{sysvan}
\end{eqnarray}
\end{subequations}
with
\begin{eqnarray}
f(a,\, \ve_{\rm vdw}) &=&  \frac{24 W_{\rm vdw}\ve_{\rm vdw}}{3 -
\ve_{\rm vdw}} - 9 \ve_{\rm vdw}^2 + \ve_{\rm vdw} + 2  \lambda_{\rm
rad} V_0^{4/3}/a^4 \nonumber \\
&+& (3W + 1)\left( \frac{A}{1 + W}+\lambda_{\rm mchap}
V_0^{(1+W)(1+\alpha)}/a^{3(1+W)(1+\alpha)}\right)^{1/(1+\alpha)}
\label{fright}\\ &-& 3 A /\left( \frac{A}{1 + W}+\lambda_{\rm mchap}
V_0^{(1+W)(1+\alpha)}/a^{3(1+W)(1+\alpha)}\right)^{\alpha/(1+\alpha)}.
\nonumber
\end{eqnarray}
In what follows, we solve the foregoing system numerically. Our goal
is here pure pedagogical, so we use possible simple values for
problem parameters setting $\kappa = 1$, $V_0 = 1$, $\lambda_{\rm
rad} = 1$, $\lambda_{\rm mchap} = 1$, $W_{\rm vdw} = 1/2$, $\alpha =
1/2$, $W = -1/2$, and $A = 1$. For initial values we use $H (0) =
0.1$, $a(0) = 0.9$ and $\ve_{\rm vdw} (0) = 0.9$. In Fig. \ref{a3n1}
we have illustrated the evolution of metric function when the
Universe is filled with Van-der-Waals gas, radiation and modified
Chaplygin gas. Evolution of corresponding Hubble parameter is given
in Fig. \ref{Hubble3n1}. The Figs. \ref{endenvdw}, \ref{pressvdw}
and \ref{eosvdw}, show the evolution of energy density, pressure and
EoS parameter of the Van-der-Waals gas, respectively. As one sees,
the Van-der-Waals gas possesses negative pressure at the early stage
which gives rise to initial inflation.

\section{Conclusion}

Within the scope of isotropic FRW cosmological model the role of
nonlinear spinor field in the evolution of the Universe is studied.
It is found that unlike in anisotropic cosmological models in the
present case the spinor field does not possess nontrivial
non-diagonal components of energy-momentum tensor. As a result,
there is no need to impose any additional restriction on either the
spinor field or on the metric function or on both of them,  as it
takes place  in case of anisotropic models. The spinor description
of different matter was given and evolutions of the Universe
corresponding to these sources are illustrated. In the framework of
a three fluid system we have shown why the spinor description of
matter is more convenient to study the evolution of the Universe
filled with multiple sources.

\vskip 7mm

\noindent {\bf Acknowledgments}\\
This work is supported in part by a joint Romanian-LIT, JINR, Dubna
Research Project, theme no. 05-6-1119-2014/2016.

Taking the opportunity I would like to thank the reviewer for
several suggestions to improve the manuscript.

\end{document}